\documentclass[twocolumn,showpacs,showkeys,preprintnumbers,amsmath,amssymb]{revtex4}

\usepackage{graphicx}
\usepackage{dcolumn}
\usepackage{bm}

\begin{document}

\title{Gravitational Waves from Periodic Three-Body Systems} 

\author{V. Dmitra\v sinovi\' c and Milovan \v Suvakov}
\affiliation{Institute of Physics Belgrade, University of Belgrade, Pregrevica 118, 11080 Beograd, Serbia }

\author{Ana Hudomal}
\affiliation{Fizi\v cki fakultet, University of Belgrade, Studentski Trg 12, 11000 Belgrade, Serbia}

\begin{abstract}
Three bodies moving in a periodic orbit under the influence of Newtonian gravity ought to emit gravitational waves. 
We have calculated the gravitational radiation quadrupolar waveforms and the corresponding 
luminosities for the 13+11 recently discovered three-body periodic orbits in Newtonian gravity. 
These waves clearly allow one to 
distinguish between their sources: all 13+11 orbits have different waveforms and their luminosities
(evaluated at the same orbit energy and body mass) vary by up to 13 orders of magnitude in the mean, 
and up to 20 orders of magnitude for the peak values. 

\end{abstract}

\maketitle

Direct detection of gravitational waves \cite{Misner:1973ad,Lightman1975} ought to come about in the 
foreseeable future, due to the substantial effort made at the operational and/or pending detectors. 
One of the most promising candidates for astrophysical sources of gravitational waves are 
the coalescing, i.e., inspiraling and finally merging binary compact stars \cite{Cutler:2002me,Cutler:1992tc}. 
Binary coalescence is the only source for which there is a clear prediction of the signal and 
an estimate of the detection distance limit, as
general relativists have completed numerical simulations of mergers of compact 
binaries, such as neutron stars and/or black holes, Refs. \cite{Pretorius:2005ad,Baker:2006ad,Campanelli:2006ad}. 

Slowly changing, quasiperiodic two-body orbits are weak sources of 
gravitational radiation, Refs. \cite{Peters:1963ad,Peters:1964ad}---only accelerated collapse leads to an increase in energy loss. 
The major part of the emitted energy in a binary coalescence
comes from the final merger of two neutron stars, or black holes, that 
produces an intense burst of gravitational radiation. 
Of course, such mergers are one-off events, never to be repeated in the same system, 
so their detection is subject to their (poorly known) 
distribution in our Galaxy. It is therefore interesting to look for periodic sources 
of intense gravitational radiation. 

There is now a growing interest in three-body systems as astrophysical sources of 
gravitational waves, Refs. \cite{Chiba:2006ad,Torigoe:2009bw,Asada:2009qs}. These early works 
did not find a substantial increase in the luminosity (emitted power) from representative 
three-body orbits belonging to three families that were known at the time, Refs. 
\cite{Broucke1975,Lagrange1772,Broucke1975b,Hadjidemetriou1975a,Hadjidemetriou1975b,Hadjidemetriou1975c,Henon1976,Henon1977,Moore1993,Simo2002}, 
over the luminosity from a comparable periodic two-body system \footnote[23] {The question of 
distinguishability between various three-body and two-body sources' of gravitational 
radiation was also raised in Ref. \cite{Torigoe:2009bw}.}. 
The luminosity  
of a (quadrupolar) gravitational wave is proportional to the square of the third time 
derivative of the quadrupole moment, see Refs. \cite{Peters:1963ad,Peters:1964ad}, 
which, in turn, is sensitive to close approaches of two bodies in a periodic orbit 
\footnote[24]{The proximity to a two-body collision can be defined 
mathematically by using the so-called hyperspherical variables, and the shape-sphere variables, 
in particular; see Refs. \cite{Suvakov:2013,Suvakov:2014a}. Of course,
it is not just the proximity to the two-body collision point that is driving this surge of emitted
power, but also the accompanying increase in the velocities, accelerations, and third derivatives
of the relative positions; see the text below.}.
Thus, getting as close as possible to a two-body collision without 
actually being involved in one, is a desirable property of the radiating system. 

Recently 13 new distinct periodic orbits belonging to 12 (new) families 
have been discovered in Ref. \cite{Suvakov:2013}, as well as 11 ``satellite orbits''
in the figure-eight family \cite{Suvakov:2013b}. Some of these 
three-body orbits pass very close to binary collisions and yet avoid them, so they are  
natural candidates 
for periodic sources of intense gravitational radiation.

\begin{figure}[t]
\includegraphics[width=1.\columnwidth]{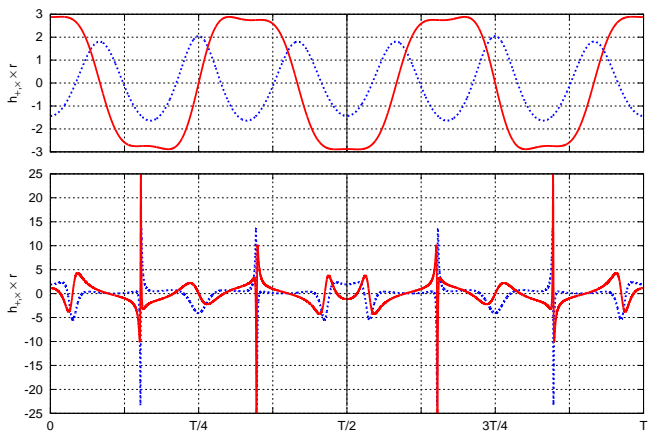}
\caption{The gravitational radiation quadrupolar waveforms $h_{+,\times} \times r$ as 
functions of the elapsed time $t$ in units of the period $T$, for two periodic three-body orbits (in units of $G m/c^2$; we have set $G=m=c=1$
throughout this Letter) 
and $r$ is the radial distance from the source to the observer. 
Dotted (blue) and solid (red) curves denote the + and $\times$ modes, respectively.
Top: Simo's figure eight, Ref. \cite{Simo2002}; and bottom: orbit I.B.1 Moth I. Note the symmetry of 
these two graphs under the (time-)reflection about the orbits' midpoint $T/2$ during one period $T$.}
\label{fig:waveforms}
\end{figure}

\begin{center}
\begin{figure}[t]
\includegraphics[width=1.\textwidth]{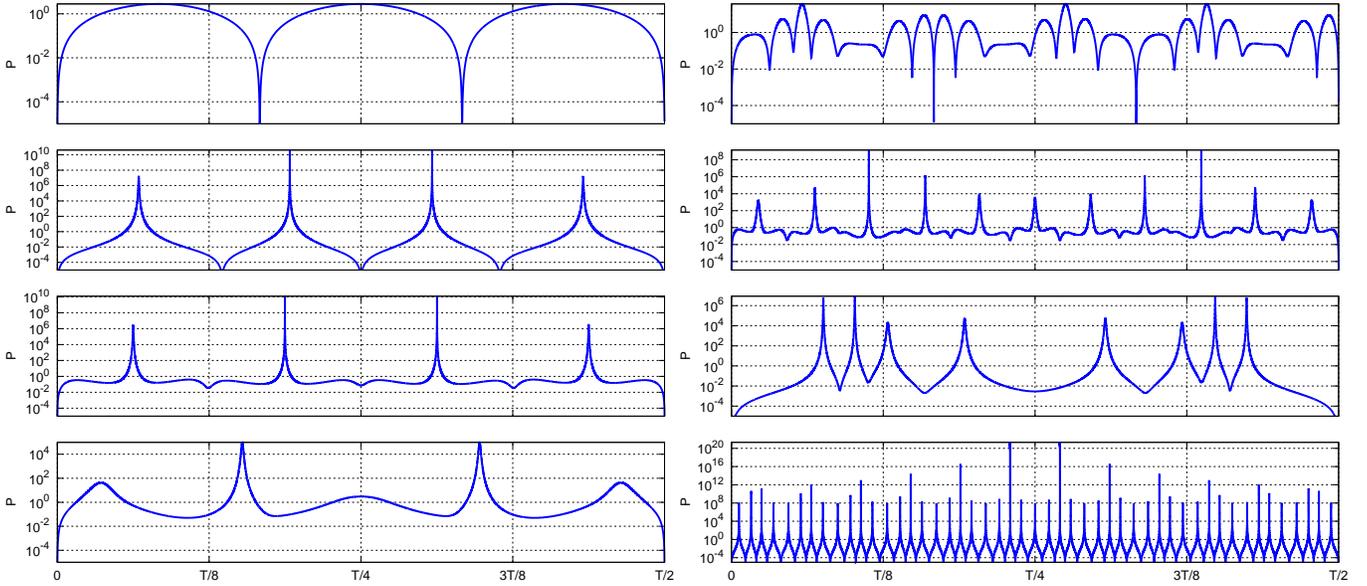}
\caption{The instantaneous (time unaveraged) luminosity $P$ of quadrupolar gravitational radiation 
emitted from periodic three-body orbits as a function of the elapsed time $t$ in units of the period $T$.
Note the logarithmic scale for the luminosity $P$ ($y$ axis). 
Top left: Moore's figure eight; second from top left: I.A.2 butterfly II; third from top left: 
II.B.7 dragonfly; bottom left: I.B.1 moth I; top right: $(M8)^7$; second from top right: 
I.A.3 bumblebee; second from bottom right: I.B.5 goggles; bottom right: II.B.6 butterfly IV.}
\label{fig:power}
\end{figure}
\end{center}

In this Letter we present our calculations of quadrupolar waveforms, 
Fig. \ref{fig:waveforms}, and of luminosities, see Table \ref{tab:power} and Fig. \ref{fig:power} 
of gravitational radiation emitted by the 13+11 recently discovered periodic three-body 
gravitating orbits, Refs. \cite{Suvakov:2013,Suvakov:2013b}.
We have also calculated waveforms of all published Broucke-Hadjidemetriou-Henon (BHH) orbits
\cite{Broucke1975,Broucke1975b,Hadjidemetriou1975a,Hadjidemetriou1975b,Hadjidemetriou1975c,Henon1976,Henon1977}, 
which we omit from this Letter for the sake of brevity, and because they are closely
related to Henon's ``criss-cross'' one, studied in Ref. \cite{Torigoe:2009bw}. 
The waves of the 13+11 new orbits show clear distinctions in form and luminosity,  
thus ensuring that they would be distinguishable (provided their signals are strong enough to be detected).

\begin{table*}[t]
\caption{Initial conditions and periods of three-body orbits.
${\dot x}_{1}(0)$, ${\dot y}_{1}(0)$ are the first particle's initial 
velocities in the $x$ and $y$ directions, respectively, 
$T$ is the period of the (rescaled) orbit
to normalized energy $E=-1/2$, $\Theta$ is the rotation angle (in radians) and 
$\langle P \rangle$ is the mean luminosity (power) of the waves emitted during one period. 
Other two particles' initial conditions are specified by these two parameters, 
as follows: $x_1(0)=-x_2(0)=-\lambda$, $x_3(0)=0$, $y_1(0)=y_2(0)=y_3(0)=0$,  
$\dot x_2(0)=\dot x_1(0)$, $\dot x_3(0)=-2\dot x_1(0)$, $\dot y_2(0)=\dot y_1(0)$,
$\dot y_3(0)=-2\dot y_1(0)$.  The Newtonian coupling constant $G$ is taken as $G=1$
and the masses are equal $m_{1,2,3}=1$.}

\begin{tabular}{lcccccc}

\hline \hline 
Name & 
\multicolumn{1}{c}{${\dot x}_{1}(0)$} & 
\multicolumn{1}{c}{${\dot y}_{1}(0)$} & 
\multicolumn{1}{c}{$\lambda$} & 
\multicolumn{1}{c}{${\rm T}$} & 
\multicolumn{1}{c}{$\Theta ({\rm rad})$} &
\multicolumn{1}{c}{$\langle P \rangle$} \\

\hline
Moore's figure eight & 0.216 343 & 0.332 029
& 2.574 29 & 26.128 & 0.245 57 & {$1.35  \times 10^0$} \\
Simo's figure eight & 0.211 139 & 0.333 568
& 2.583 87 & 26.127 & 0.277 32 & $1.36  \times 10^0$ \\
$({\rm M}8)^7$ & 0.147 262 & 0.297 709
& 3.008 60 & 182.873 & 0.269 21 & $2.46  \times 10^0$ \\
I.A.1 butterfly I & 0.147 307 & 0.060 243 
& 4.340 39 & 56.378 & 0.034 78 & $1.35 \times 10^5$ \\ 
I.A.2 butterfly II & 0.196 076 & 0.048 69 
& 4.016 39 & 56.375 & 0.066 21 & $5.52 \times 10^6$ \\ 
I.A.3 bumblebee & 0.111 581 & 0.355 545
& 2.727 51 & 286.192 & -1.090 4 & $1.01 \times 10^5$  \\ 
I.B.1 moth I & 0.279 332 & 0.238 203
& 2.764 56 & 68.464 & 0.899 49 & $5.25 \times 10^2$ \\ 
I.B.2 moth II & 0.271 747 & 0.280 288
& 2.611 72 & 121.006 & 1.138 78 & $1.87 \times 10^3$ \\ 
I.B.3 butterfly III & 0.211 210 & 0.119 761
& 3.693 54 & 98.435 & 0.170 35 & $3.53 \times 10^5$ \\ 
I.B.4 moth III & 0.212 259 & 0.208 893
& 3.263 41 & 152.330 & 0.503 01 & $7.48 \times 10^5$ \\ 
I.B.5 goggles & 0.037 785 & 0.058 010
& 4.860 23 & 112.129 & -0.406 17 & $1.33 \times 10^4$ \\ 
I.B.6 butterfly IV & 0.170 296 & 0.038 591
& 4.226 76 & 690.632 & 0.038 484 & $1.23 \times 10^{13}$ \\ 
I.B.7 dragonfly & 0.047 479 & 0.346 935
& 2.880 67 & 104.005 & -0.406 199 & $1.25 \times 10^6$  \\ 
II.B.1 yarn & 0.361 396 & 0.225 728
& 2.393 07 & 205.469 & -1.015 61 & $2.33 \times 10^6$ \\ 
II.C.2a yin-yang I & 0.304 003 & 0.180 257
& 2.858 02 & 83.727 & 0.659 242 & $1.31 \times 10^5$ \\ 
II.C.2b yin-yang I & 0.143 554 & 0.166 156
& 3.878 10 & 83.727 & -0.020 338 & $1.31 \times 10^5$ \\ 
II.C.3a yin-yang II & 0.229 355 & 0.181 764
& 3.302 84 & 334.877 & 0.472 891 & $7.19 \times 10^{10}$ \\ 
II.C.3b yin-yang II & 0.227 451 & 0.170 639
& 3.366 76 & 334.872 & 0.254 995 & $7.19 \times 10^{10}$ \\ 
\hline
\hline
\end{tabular}
\label{tab:power}
\end{table*}

We consider systems of three equal massive particles moving periodically in a plane 
under the influence of Newtonian gravity.
The quadrupole moment $I_{ij}$ of three bodies with equal masses $m_n =m$, ($n=1,2,3$) is expressed as 
$I_{ij} = \sum_{n=1}^{3} m ~x^{i}_{n} x^{j}_{n}$ , where $x_n^{i}$ is the location 
of $n$th body, and the spatial dimension indices $i$ and $j$ run 
from 1 to 3 (with $x^1 = x$, $x^2 = y$, $x^3 = z$). The reduced quadrupole $Q_{ij}$ is defined as 
$Q_{i j} = I_{i j} - \frac13 \delta_{i j} \sum_{k=1}^3 I_{k k}$. 
The gravitational waveforms denoted by $h_{ij}^{TT}$ are, asymptotically,
\begin{eqnarray}
\label{eq:0}
h^{TT}_{ij} &=& \frac{2G}{r c^4} \frac{d^2 {Q}_{ij}}{dt^2} 
+ {\cal O}\left(\frac{1}{r^2}\right),\
\end{eqnarray}
where $r$ is the distance from the source, Refs. \cite{Peters:1963ad,Peters:1964ad}.
Here, $TT$ means (i) transverse ($\sum_{i=1}^3 h^{TT}_{ij} {\hat n}^i = 0$) and (ii) traceless
($\sum_{i=1}^3 h^{TT}_{ii} = 0$), where ${\hat n}_i$ denotes the unit vector of the 
gravitational wave's direction of propagation. The two independent waveforms $h_{+,\times}$ 
of a quadrupolar gravitational wave propagating along  
the $z$ axis, Refs. \cite{Peters:1963ad,Peters:1964ad} can be expressed as 
\begin{eqnarray}
\label{eq:1}
h_{+} &=& 
\frac{2G}{c^4r} \sum_{i=1}^{3}{m_i(\dot{x_i}^2+x_i\ddot{x_i}-\dot{y_i}^2-y_i\ddot{y_i})}, \\
\label{eq:2}
h_{\times} &=& \frac{2G}{c^4r}\sum_{i=1}^{3}{m_i(\ddot{x_i}y_i+2\dot{x_i}\dot{y_i}+x_i\ddot{y_i})},\
\end{eqnarray}
where $r$ denotes the distance from the source to the observer. We set the units of $G=c=m=1$
throughout this Letter.

Here the coordinate axes $x$ and $y$ are chosen so that they 
coincide with the orbits' two (reflection) symmetry axes, when they exist, 
i.e., when the orbits are from class I, as defined in Ref. \cite{Suvakov:2013}. 
Otherwise, e.g., when only a single point reflection symmetry exists, as in class II orbits, 
the $x$, $y$ axes are taken to be the eigenvectors of the moment-of-inertia tensor. 
The rotation angle necessary for each orbit to be aligned with these two axes 
is given in Table \ref{tab:power}
\footnote[27]{When the orbit passes through the Euler point twice, such as in the yin-yang 
orbits, there are two different sets of initial conditions, and, consequently, two different
rotation angles---we indicate exactly which one of the two solutions is taken. 
The total energy has been scaled to $E=-1/2$ for all solutions, so as to provide a 
meaningful comparison of peak amplitudes and luminosities.}.

The first gravitational radiation waveforms for periodic three-body systems were studied 
in Refs. \cite{Torigoe:2009bw,Chiba:2006ad,Asada:2009qs}. 
They calculated the quadrupole radiation waveforms for three periodic orbits of the following 
three-equal-mass systems: 
(i) of the Lagrange ``equilateral triangle'' orbit \cite{Lagrange1772}, 
(ii) of Henon's ``criss-cross'' \cite{Henon1976}, and 
(iii) of Moore's ``figure eight'' \cite{Moore1993}. 
These three orbits are characteristic representatives of the (only) three families 
of periodic three-body orbits 
known at the time. Reference \cite{Torigoe:2009bw} found distinct gravitational waveforms for each 
of the three families, thus suggesting that one might be able 
to distinguish between different three-body systems as sources of gravity waves 
by looking at their waveforms \footnote[28]{A more detailed study of the waveforms emanating 
from the Lagrangian three-body orbit can be found in Ref. \cite{Asada:2009qs}.}.

In the meantime 13+11 new orbits belonging to 12 new families 
have been found, Refs. \cite{Suvakov:2013,Suvakov:2013b}.
The families of three-body orbits can be characterized by their topological
properties viz. the conjugacy classes of the fundamental group, in this case, the free 
group on two letters (${a,b}$), Ref. \cite{Suvakov:2014a}. 
The free group element tells us the number of times the system's trajectory on the shape sphere 
passes around one or another (prechosen) two-body collision point within one period.
Every time the system is close to a two-body collision the (relative) velocities, accelerations,
and the third derivatives of relative coordinates increase, so that the luminosity of gravitational 
radiation also increases; i.e., there is a burst of gravitational radiation. This argument can 
be made more quantitative by appealing to two-body results of Ref. \cite{Peters:1963ad},
as is shown in footnote [32].

We show the gravitational radiation waveforms $h_{+,\times}$ in Fig. \ref{fig:waveforms}, 
emitted by three massive bodies moving according to the orbits from
Refs. \cite{Suvakov:2013,Suvakov:2013b} belonging to these families, where Eqs. (\ref{eq:1}) 
and (\ref{eq:2}) are used as the definitions of the two waveforms.

First, we note that all of the calculated 
three-body orbits' waveforms are distinct \footnote[31]{We do not show these waveforms here, 
except for the two in Fig. \ref{fig:waveforms}, for brevity's sake, and because 
many are fairly similar to the second waveform in Fig. \ref{fig:waveforms}---regular 
sequences of spikes.}, thus answering (in the positive) the question 
about their distinguishability posed in Ref. \cite{Torigoe:2009bw}. 
In Fig. \ref{fig:waveforms} we also show the gravitational waveform of one ``old'' orbit: 
Simo's figure eight, (discovered in 2002) belonging to the figure-eight family. Simo's figure eight is an important example, as it is virtually indistinguishable from Moore's one, 
and yet the two have distinct gravitational waveforms, see our Fig. \ref{fig:waveforms} 
and Fig. 2 in Ref. \cite{Torigoe:2009bw}. That is so because 
these two figure-eight solutions have distinct time dependences of the hyperradius $R$, where
$R^2 \sim (1/m) \delta_{i j} \sum_{k=1}^3 I_{k k}$,  
so that the two orbits have different quadrupolar waveforms.

Note, moreover, the symmetry of the waveforms in Fig. \ref{fig:waveforms} with respect 
to reflections of time about the midpoint of the period $T/2$: this is a consequence of 
the special subset of initial conditions (vanishing angular momentum and passage through 
the Euler point on the shape sphere) that we used. There are periodic three-body orbits, 
such as those from the BHH family, that do not have this symmetry.

The gravitational waveforms' maxima range from 20 to 50 000 in our units, with the energy fixed 
at $E=-1/2$. This large range 
of maximal amplitudes is due to the differences in the proximity of the approach to two-body collisions 
in the corresponding orbits. One can explicitly check that the 
bursts of gravitational radiation during one period correspond  
to close two-body approaches.

As stated above, the (negative) mean power loss $\left\langle dE/dt \right\rangle$ of the 
three-body system, or the (positive) mean luminosity (emitted power) of quadrupolar gravitational 
radiation $\left\langle P \right\rangle$, averaged over one period, 
is proportional to the square of the third time derivative of the (reduced) quadrupole moment
$Q^{(3)}_{jk}$, $\left\langle dE/dt \right\rangle = - \left\langle P \right\rangle = - 
\frac15 (G/c^5) \sum_{j,k=1}^3 \left\langle Q^{(5)}_{jk} {\dot Q}_{jk} \right\rangle 
= - \frac15 (G/c^5) \sum_{j,k=1}^3 \left\langle  Q^{(3)}_{jk} {Q}^{(3)}_{jk} \right\rangle$,
(for an original derivation see Refs. \cite{Peters:1963ad,Peters:1964ad}, for pedagogical ones, see
Refs. \cite{Misner:1973ad,Lightman1975}).
But, $Q^{(3)}_{jk}$ are proportional to the first time derivatives of the gravitational waveforms
$Q^{(3)}_{jk} = (d/dt) Q^{(2)}_{jk} \propto (d/dt) h_{+,\times}$. The peak amplitudes
of gravitational waveforms $h_{+,\times}$, in turn, grow  in the vicinity of two-body collisions
\footnote[30]{The following argument was suggested by one of the referees:
If a section of the trajectory of two bodies (within a three-body system) that approach a two-body 
collision can be approximated by an ellipse, then the luminosity $P$ is proportional to 
$P \sim (1-e^2)^{-7/2}$, see Eq. (5.4) in Ref. \cite{Peters:1964ad}, where $e$ is the eccentricity 
of the ellipse. Therefore, $P$ grows without bounds as $e \to 1$, i.e., as the orbit approaches a 
two-body collision.}, 
which explains the burst of gravitational radiation as one approaches a two-body collision
point.

The mean and instantaneous luminosities, expressed in our units, of these orbits, normalized to $E=-1/2$, 
are shown in Table \ref{tab:power} and Fig. \ref{fig:power}, respectively. 
Note that in Table \ref{tab:power} we show only three of the 11 orbits belonging to the figure-eight family: 
Moore's, Simo's, and the stable choreography $({\rm M}8)^7$; they have all the same order of magnitude of 
the mean luminosity \footnote[32]{Note that the figure-eight family members have, on the average, the lowest 
luminosity among the orbits considered here.}, whereas the butterfly I and butterfly II orbits, 
which belong to the same topological family, have mean luminosities that differ by more than 
a factor of 40.

Generally, the mean 
luminosities of these 24 orbits cover 13 orders of magnitude, ranging from 1.35 (Moore's figure eight) 
to $1.23 \times 10^{13}$ (I.B.6 butterfly IV) in our units; see Table \ref{tab:power}. 
The peak instantaneous luminosities have an even larger range: 20 orders of magnitude; 
see Fig. \ref{fig:power}. Here, the symmetric form of the instantaneous (time unaveraged) power 
$P = \frac15 (G/c^5) \sum_{j,k=1}^3 Q^{(3)}_{jk} {Q}^{(3)}_{jk}$ was used.
This gives us hope that at least some of these three-body periodic orbits can, perhaps, 
lead to detectable gravitational radiation signals. 

It is a different question if some or all of these sources of gravitational radiation 
would be observable by the present-day and the soon-to-be-built gravitational wave detectors: that
strongly depends on the absolute values of the masses, velocities, and the average distances 
between the three celestial bodies involved, as well as on the distribution of such sources in our Galaxy. 

Moreover, note that all of the newly found and analyzed three-body orbits have zero angular momentum,
and many of them are unstable. It is well known 
\cite{Hadjidemetriou1975a,Hadjidemetriou1975b,Hadjidemetriou1975c,Henon1976,Henon1977} that by changing
the angular-momentum within the same family of three-body orbits, the stability of an orbit changes as well.
So, it may happen that a previously stable orbit turns into an unstable one, and vice versa. 
For this reason it should be clear that a careful study of 
gravitational-radiation-induced energy- and angular-momentum 
dissipation is necessary for these orbits \footnote[33]{We plan to do such a study, which cannot be 
completed, however, without an extension of each orbit to a family of orbits with nonvanishing 
angular momenta. So far, only the BHH family has been extended in such a way, but even that one 
case is not complete \cite{Jankovic:2013}.}. 
Moreover, if realistic results are to be obtained, post-Newtonian approximations will have 
to be applied in the future. Such relativistic corrections are most important at large velocities, 
i.e., precisely near close approaches that are so crucial for large gravitational radiation. Thus, 
the present Letter is meant only to highlight the possibilities in this field, and should be 
viewed as an invitation to join in the more realistic future studies. 


V. D. and M. \v S. were financially supported  by the Serbian Ministry of Science and
Technological Development under Grants No. OI 171037 and No. III 41011. A. H. 
was supported by the City of Belgrade studentship
(Gradska stipendija grada Beograda) during the 
year 2012--2013, and was a 
recipient of the ``Dositeja'' stipend for the year 2013--2014, from the Fund for
Young Talents (Fond za mlade talente - stipendija ``Dositeja'') 
of the Serbian Ministry for Youth and Sport.

\onecolumngrid
\twocolumngrid

\end{document}